\documentclass[11pt,twoside]{article}

\usepackage{asp2004}
\usepackage{epsf}
\usepackage{psfig}
\usepackage{lscape}

\begin{document}
\title{A Short and Personal History of the {\it Spitzer Space Telescope}}
\author{Michael Werner}   
\affil{JPL/Caltech}    

\begin{abstract} 

The {\em Spitzer Space Telescope}, born as the {\em Shuttle Infrared
Telescope Facility} ({\em SIRTF}) and later the {\em Space Infrared
Telescope Facility} (still {\em SIRTF}), was under discussion and
development within NASA and the scientific community for more than
30 years prior to its launch in 2003.  This brief history chronicles
a few of the highlights and the lowlights of those 30 years from the
authors personal perspective. A much more comprehensive history of {\em
SIRTF}/{\em Spitzer} has been written by George Rieke (2006).

\end{abstract}



\section{Pre-1983}

{\em SIRTF} first appeared on the scene in 1971 in response to a search
for payloads for the then recently-proposed Space Shuttle.  To quote from
Witteborn and Manning (1988), ``The preliminary plan for a far infrared
‘Cooled Telescope for Sortie Mode Shuttle’ (Figure~1) was prepared at
NASA's Ames Research Center (ARC) and presented to NASA Headquarters in
May 1971.''  The following year, an infrared panel convened as part of
a Space Shuttle Sortie Workshop recommended development of a one-meter
class cooled telescope for flight aboard the Shuttle, which at that
time was expected to provided frequent flight opportunities and missions
lasting up to 30 days.

Work on a shuttle-based concept –-- which eventually became known
as the {\em Shuttle Infrared Telescope Facility} –-- continued
from 1971 through 1984 under ARC leadership.  Community-based groups
defined scientific objectives, scientific requirements, and strawman
instrument concepts for a cooled, one-meter class telescope. The present
author's involvement with {\em SIRTF} began in 1977 with membership
in one of these groups. Industrial studies were carried out by aerospace
companies including Martin Marrietta, Hughes Aircraft, and Perkin Elmer.
At the same time, fed by the results of these studies, the {\em SIRTF}
concept was gaining traction within the wider scientific community.
Study groups chartered by the National Academy of Sciences and the Space
Science Board recognized the power of a cooled infrared telescope in
space and recommended continued NASA investment in science, system,
and mission studies and, critically, in technology development.

The decadal astronomy review which set priorities for astronomy in the
1980’s, called the Field report in recognition of George Field who
chaired the review process, included the Shuttle-attached {\em SIRTF}
in a category called ``approved and  continuing programs ... from which
the recommendations of the [Field] Committee proceed.''  Significantly,
the Field Committee also recommended study of the ways by which {\em
SIRTF} could become a ``... long duration observatory ...''   Later in
the decade, when {\em SIRTF} had evolved from the shuttle-attached to
the free-flying implementation, these words did not carry enough weight
to allow {\em SIRTF} to start development before the Committee's
highest priority, then known as {\em AXAF}, now the {\em Chandra X-ray
Observatory}.

The subsequent evolution of {\em Spitzer}, from the shuttle-borne
concept through a number of free-flying versions (see Figure~2) into the
compact and elegant observatory now operating, involved numerous setbacks
and delays.  However, NASA's investment in relevant technologies ---
detectors, cryogenics, and optics --- continued through the years.
The detector array development program, started in 1978 under the
direction of Craig McCreight at ARC merits special mention.

\begin{figure}
\plotone{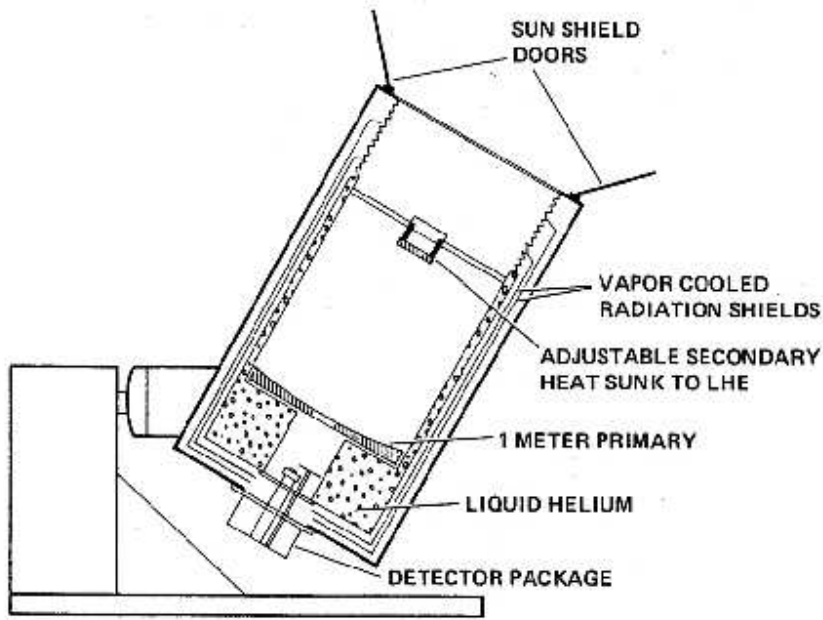}
\caption{Preliminary concept for a far infrared ``liquid helium cooled
telescope for sortie mode shuttle,'' 1971.  The original drawing notes
that helium gas vents, a removable vacuum cover, and a ``finder scope
with TV'' are not shown.}
\label{fig1}
\end{figure}

\section{1983 -- 1984}

NASA Announcement of Opportunity OSSA-1-83 for the {\em Shuttle Infrared
Telescope Facility}, released May 13, 1983, envisaged {\em SIRTF} as
``... an attached Shuttle mission with an evolving scientific payload'',
but hinted at ``... a probable transition to a more extended mode of
operation, possibly in association with a future space platform or
space station.''  The AO solicited proposals for investigations which
required development of focal plane instruments as well as for individual
investigations in the Facility Scientist and Interdisciplinary Scientist
categories.  The first shuttle flight of {\em SIRTF} was envisioned
for ``... about 1990 with the second flight ... approximately one year
after the first flight.''  The release of the AO was a major turning
point for {\em SIRTF}.  The words in the Field report certainly helped,
but most of the credit for this milestone goes to Nancy Boggess and
others at NASA Headquarters who recognized the importance of following up
the spectacular success of the recently-launched {\em IRAS} satellite
({\em cf.} Rieke, 2006).  A similar response to the success of {\em IRAS}
within the European community led to ESA's very successful {\em Infrared
Space Observatory} ({\em ISO}) mission, launched in 1995.

\begin{figure}
\plotone{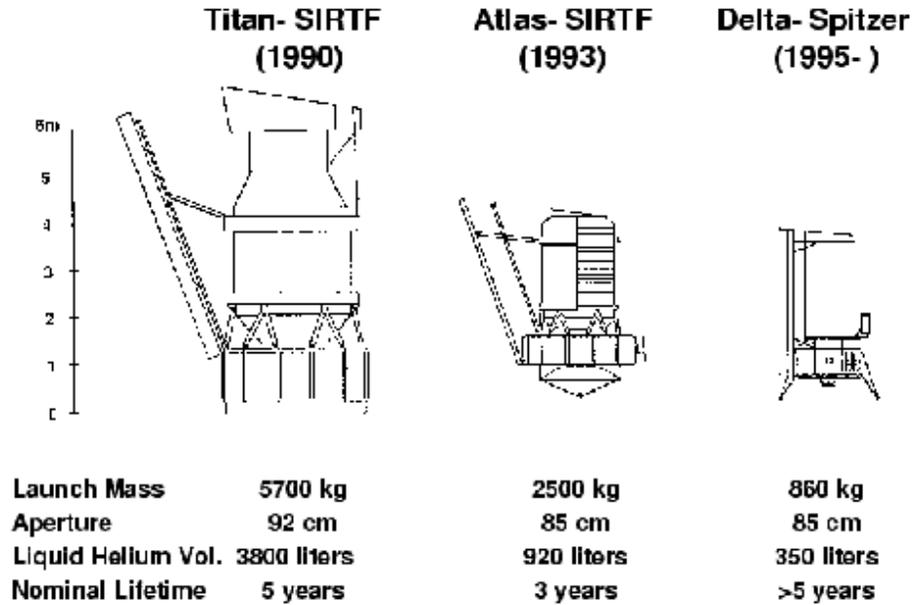}
\caption{The evolution of the free-flying {\em Spitzer}.}
\label{fig2}
\end{figure}

There were a few odd clauses in the AO which, fortunately, were rendered
inoperative by the move to the free flyer.  In particular, an obscure
paragraph states that there might be reason to fly one or two {\em SIRTF}
science instrument team members on the Shuttle as Payload Specialists to
operate {\em SIRTF}.  It goes on to suggest that  ``This option ... can be
exercised if the {\em SIRTF} Science Working Group (SWG) can ... produce
suitably qualified candidate scientists.''  Fortunately, this is one
issue which the SWG was never forced to confront.  The AO also promised
an overselection of instruments followed by a subsequent down-select,
but this option was abandoned in favor of more robust support for the
three selected instruments.

Payload Specialists aside, the Shuttle-based {\em SIRTF} was, in fact,
rendered dead on arrival by the success of {\em IRAS}, launched on January
26, 1983.   At the {\em SIRTF} pre-proposal briefing on July 11, 1983,
NASA Program Scientist Nancy Boggess reported that the early data from
{\em IRAS} was transforming {\em SIRTF} into a free-flyer. The critical
data included both scientific results revealing the richness of the
infrared sky, and the technical success of {\em IRAS} in demonstrating the
cryogenic and detector performance required for a {\em SIRTF} free flyer.
At the same time, Shuttle-based experiments were raising concerns about
the particulate and radiative cleanliness of the Shuttle environment.

NASA therefore changed the rules in midstream and issued an amendment
to the Announcement of Opportunity on September 12, 1983, asking for
proposals which laid out a science program suitable for a hypothetical
long-life mission as well as one suitable for two two-week long shuttle
missions.  The long-life mission duration was set at one year, and the
selected orbit was a polar, sun-synchronous orbit at 900 km altitude,
just like {\em IRAS'}.

The Amendment set the due date for proposals as December 5, 1983.  On May
3, 1984, the Ames team –-- with the support of the scientific community
--- formally recommended to NASA that {\em SIRTF} be developed as a
free-flying observatory.  The relevant NASA officials --– Associate
Administrator Burton Edelson and Administrator James Beggs –--
accepted and concurred in this recommendation and, on June 20, 1984,
the {\em Space Infrared Telescope Facility} came into being.

\section{1984 -- 1989}

In June, 1984, a NASA press release and letters to the proposers announced
the results of the AO selection.  Three teams, headed by Principal
Investigators Giovanni Fazio, Jim Houck, and George Rieke, were selected
to develop instruments – which have evolved into the current day IRAC,
IRS, and MIPS. Frank Low was selected as Facility Scientist and Mike Jura
and Ned Wright as Interdisciplinary Scientists.  NASA Scientists Nancy
Boggess (Program Scientist), myself (Project Scientist/SWG Chair) and
Fred Witteborn (Deputy Project Scientist/Deputy SWG Chair) completed the
SWG. The SWG has grown since 1984 with the addition of Dale Cruikshank,
Bob Gehrz, Charles Lawrence, Marcia Rieke, and Tom Roellig.  Tom Soifer,
Director of the {\em Spitzer} Science Center (SSC), is an {\em ex
officio} member, and Fred Witteborn, who pioneered the {\em SIRTF}
concept in the early 1970’s, left the SWG when {\em SIRTF} moved from
Ames to JPL.  It is  noteworthy that, more than 20 years later, all six
initially-selected SWG members as well as the present author remain very
active in the {\em Spitzer} science programs.  Along the way we were
fortunate to work with a number of extremely capable Project Managers,
most notably Larry Simmons, Dave Gallagher, and Bill Irace at JPL, but
it was the scientists who provided the inspiration and continuity which
brought {\em Spitzer} to fruition.

\begin{figure}
\plotone{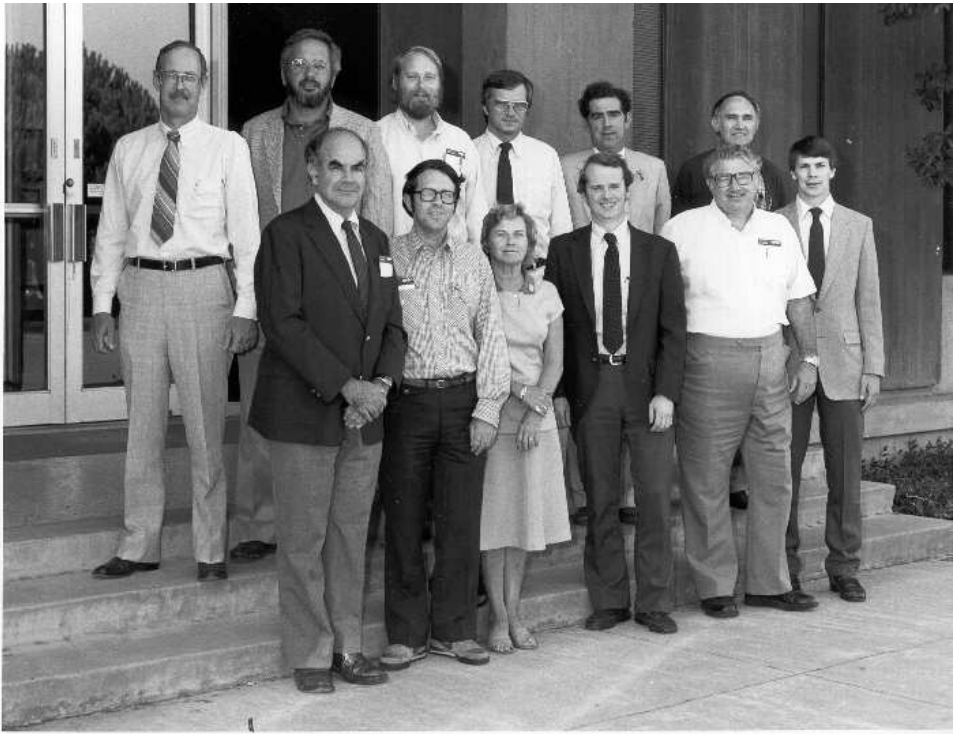}
\caption{The first meeting of the {\em Spitzer} (then {\em SIRTF}) Science
Working Group was held at the NASA Ames Research Center, September, 1984.
Rear row (L to R): George Newton, NASA Program Manager; Dan Gezari,
NASA-Goddard; Ned Wright and Mike Jura, UCLA; Mike Werner and Fred
Witteborn, NASA-Ames; Giovanni Fazio, SAO; George Rieke, Arizona; Nancy
Boggess, NASA Program Scientist; Jim Houck, Cornell; Frank Low, Arizona;
Terry Herter, Cornell.}
\label{fig3}
\end{figure}

The SWG had the first of its fifty-some face-to-face meetings at the
Ames Research Center on September 12-14, 1984 (Figure~3).  During the
period from 1984 to 1989 the SWG and the {\em SIRTF} Project Office at
Ames engaged in a constant process of redefinition of {\em SIRTF} driven
by a constantly changing programmatic landscape.  Throughout this period,
the critical detector technology studies continued by direct {\em SIRTF}
Project funding of the selected instrument teams, supplemented by funds
from McCreight's program, and through NASA Research and Analysis grants.
{\em SIRTF} also benefited from inclusion in 1985 in the family of Great
Observatories, which brought the {\em Compton Gamma Ray Observatory},
{\em Hubble Space Telescope}, {\em Chandra X-ray Observatory}, and {\em
SIRTF} under a common programmatic envelope.  This created a funding
line with some durability; in addition, the launch and eventual success
of {\em Compton}, {\em Hubble}, and {\em Chandra} gave this program high
visibility which enhanced our advocacy when we described {\em SIRTF} as
``completing the Great Observatories''.

A major conceptual breakthrough came starting in 1987 with the realization
that a free-flyer in High Earth Orbit, a.k.a. HEO ($\sim$ 100,000~km,
to be far above the Earth's radiation belts) would have major thermal
and operational advantages over a Low Earth Orbit (LEO) mission at the
cost of some added susceptibility to Galactic and solar cosmic rays.
This possibility surfaced as part of a study initiated at Ames in order
to maintain flexibility in the view of programmatic uncertainties in
the post-Challenger era. At the request of NASA HQ, a group of engineers
and mission designers from JPL worked with the Ames team to define this
mission, and the design evolved into a very large [and costly] concept
to be launched on a Titan-Centaur (Figure~2).  This concept formed the
basis of a presentation to NASA Associate Administrator Len Fisk on
March 24, 1989 which strongly recommended adoption of the HEO concept
for {\em SIRTF}.  {\em IRAS} veterans Gerry Neugebauer and Fred Gillett
(who had replaced Nancy Boggess as the {\em SIRTF} Program Scientist at
HQ) joined Frank Low and the rest of the SWG in presenting a compelling
scientific case for the HEO concept, while Project Manager Walt Brooks
presented the equally compelling technical rationale.   This presentation
ended with the adoption by Fisk of the HEO approach for {\em SIRTF}.

\section{1989 -- 1993}

In 1989, NASA questioned the ability of ARC to shepherd the HEO {\em
SIRTF} successfully through development.  Consequently, Charlie Pellerin,
head of NASA's Astrophysics Division, invited formal proposals from
NASA centers for the management of {\em SIRTF}.  JPL, ARC, GSFC, and
MSFC (teaming with Ames) submitted proposals, and NASA selected JPL.
{\em SIRTF} was formally moved to JPL in the late Fall of 1989, and {\em
SIRTF} activities at JPL began in earnest in 1990.  Werner moved from Ames
to JPL to continue as Project Scientist and SWG Chair; the contracts with
the instrument teams and the SWG members were also transferred to JPL,
but some ongoing optics technology work finished up under ARC leadership.
Dick Spehalski and Earl Cherniack, who had just successfully brought
{\em Galileo} to launch, were chosen to lead the JPL team.

Amongst the first opportunities faced by the new {\em SIRTF} team
was to bring {\em SIRTF} before the newly-constituted National
Academy of Sciences panel, led by John Bahcall,  which was establishing
priorities for astronomy and astrophysics for the decade of the 1990’s.
This process began with a meeting of an infrared panel of the Bahcall
committee, chaired by Fred Gillett and Jim Houck, held in Tucson on May
5, 1990.  The {\em SIRTF} concept presented there, the Titan-launched
{\em SIRTF} (Figure~2), was the most ambitious version of {\em SIRTF} ever
contemplated.  {\em SIRTF} emerged as the highest priority recommendation
of the infrared panel, and, thanks to the effective advocacy of Jim
Houck and Chas Beichman, was designated by the full Bahcall committee
as the highest priority major new initiative for the 1990’s.

Although JPL had and has a tradition of doing major projects in-house,
there was never any intention of implementing {\em SIRTF} as other than a
contracted or out-of-house program at JPL.  Consequently an RFP soliciting
proposals for Phase B (concept definition) studies of a Titan-launched
{\em SIRTF} was prepared at JPL and scheduled for release early in 1991.
However, at the last minute, due to a variety of problems with other major
missions –-- including {\em Hubble's} mirror --- Congress ordered NASA
to put the procurement on hold.  For much of the following two years the
status of {\em SIRTF} was very uncertain.  When the high gain antenna
on the {\em Galileo} spacecraft jammed and failed to open, JPL asked
Spehalski to lead the recovery, which left {\em SIRTF} without strong
project level leadership.  The {\em Galileo} and {\em Hubble} problems,
together with the loss of {\em Mars Observer} in 1993, led NASA to shy
away from \$2B plus projects like the Titan {\em SIRTF}.  In addition, Dan
Goldin took over as NASA Administrator in 1992 and started promulgating
the ``faster, better, cheaper'' approach which was also not congenial
to very large projects.  At one point, {\em SIRTF} was in such bad odor
that NASA HQ forbade us to use the name, and instead we pretended to be
working on something called the ``{\em Infrared Astronomy Mission}.''

All of this chaos notwithstanding, important work continued on what
was to become the {\em Spitzer Space Telescope}.  Firstly, the detector
technology work continued apace.  By this time, the instrument development
contractors were in place –-- GSFC for IRAC and Ball for both MIPS
and IRS –-- and all three teams were actively engaged in detector and
readout development keyed to the specific needs of their instrument
concepts. We attribute the great scientific success of {\em Spitzer}
in large part to the power of these detector arrays, which were supplied
by SBRC/Raytheon for the IRAC, by Rockwell/Boeing for IRS and MIPS, and,
additionally, by LBNL and the University of Arizona for MIPS. In addition,
in March, 1992, during the initial descoping of the Titan {\em SIRTF},
JPL Mission Engineer Johnny Kwok suggested that we consider using a
solar orbit rather than HEO for {\em SIRTF}.  The advantages of solar
orbit for {\em SIRTF} are well-documented (Kwok {\em et al.} 2004)
and will not be repeated here; suffice it to say that the solar orbit
(in its Earth-trailing version) was quickly adopted.

In March, 1993, we completed a study of the so-called ``Atlas {\em
SIRTF}'', intended to be launched on an Atlas into an Earth-trailing
solar orbit (Figure~2).  This version of {\em SIRTF}, intermediate in
scope between the Titan and eventually adopted Delta versions,  was also
intermediate in measurement functionality.  But it, too, was swept aside
by a NASA ruling that no single space mission could exceed \$500M in cost,
as well as by instructions from new Astrophysics director Dan Weedman
that we could use any launch vehicle we liked as long as it was a Delta.

\section{1993 -- A Critical Year}

On November 11, 1993, following a suggestion from Jim Houck, the SWG met
at a Ball Aerospace retreat in Broomfield, Colorado, to attempt to find
a way to move forward.  Over the past few years we had been gradually
descoping our scientific desires and instrument functionality requirements
to support the facility descopes, but this descoping was definitely
evolutionary or gradual rather than revolutionary. As a result, we were
still talking about a good fraction of the 9 distinct science themes
presented to the Bahcall committee in 1991. Midway through the first day
of the retreat, Mike Jura and George Rieke proposed instead that we focus
on a very small number of high level scientific questions. This was such
an appealing approach that we adopted it almost instantly and identified
four such questions very quickly (Table~1) which, not coincidentally,
were traceable to the Bahcall report.  Only the needs of these science
themes were to place requirements on the facility, but we reasoned that
a system optimized to study these questions would of course have very
powerful capabilities for exploration of a wide range of scientific
issues.  This list remained valid to launch, as progress achieved by
other means has increased the urgency of these scientific questions;
indeed, about 70\% of the observations in the SSC data base at launch
addressed one or another of these four themes.

\begin{table}[!ht]
\smallskip
\smallskip
\smallskip
\begin{center}
{\small
\begin{tabular}{l}
Table 1.  Defining Scientific Programs for {\em SIRTF} \\
\tableline
\tableline
\noalign{\smallskip}
Protoplanetary and Planetary Debris Disks \\
Brown Dwarfs and Super Planets \\
Ultraluminous Galaxies and Active Galactic Nuclei \\
The Early Universe \\
\noalign{\smallskip}
\tableline
\end{tabular}
}
\end{center}
\end{table}

The second morning of the Broomfield meeting, Frank Low produced a cartoon
sketch of the warm launch architecture for {\em SIRTF}.  The technical
details of this revolutionary approach are presented elsewhere (Finley
{\em et al.} 2004); suffice it to say that it uses radiative cooling
into deep space to extract most of the thermal energy from the telescope,
which is launched at ambient temperature and pressure rather than within
a vacuum-bearing cryostat, as was done for {\em IRAS} and {\em ISO}.
The instruments are launched cold, but within a much smaller cryostat
than would be needed to contain the telescope.  Thus the size of
the optical system is decoupled from the size of the cryostat; this
provided a path to future missions with large, cooled apertures which
could not reasonably be launched within a cryostat.  The warm launch
architecture is particularly well-suited to the solar orbit, in which
the spacecraft attitude relative to the sun remains fixed so that the
sun always shines on the solar panel/sun shield while the opposite side
of the telescope tube is painted black to serve as a radiator. In the
end, the warm launch architecture allowed the same size telescope and
lifetime as envisioned for the much more expensive Titan {\em SIRTF},
but at a fraction of the cost and mass.  This innovation allowed {\em
SIRTF} to become a ``poster child'' for Dan Goldin's new ``faster,
better, cheaper'' philosophy ({\em cf.} Figure~2).

A third important outcome of the Broomfield meeting grew out of a pithy
comment by Jim Houck emphasizing the inadequacy of the approach to
{\em SIRTF} advocacy which the SWG was adopting.  From this point on,
we always put advocacy –-- by which was meant advancing the cause
of {\em SIRTF} within Congress, NASA, OMB, or anywhere else we could
go –-- at the top of the agenda, and at least once a year for five
or six years Marcia Rieke organized a series of visits by {\em SIRTF}
scientists to congressional offices.  Although these visits were largely
informational and low key, there was certainly more than one occasion
over the next few years when a staffer's familiarity with {\em SIRTF}
or his/her memory of a recent visit cemented our position ``above the
line'' in a list of programs being recommended for funding.

From Broomfield on, although the path was by no means totally free of
obstacles, {\em SIRTF} began to move forward at an accelerating pace.
The first major next step –-- in November 2003 --- was the appointment by
JPL of Larry Simmons as the Project Manager.  Simmons had just seen the
upgraded {\em HST} camera WFPC-2 through to launch and brought to the
{\em SIRTF} job, in addition to outstanding interpersonal, technical and
managerial skills, both a very high degree of credibility at NASA-HQ and an
outstanding team of JPL engineers who had worked with him on WFPC-2. Our
next challenge was a review chartered by NASA Headquarters intended to
establish whether {\em SIRTF} (and the companion airborne {\em SOFIA}
project) were sufficiently compelling following the extensive rescoping
which each had undergone to continue to command the high ratings they had
been given by the Bahcall committee.  This review was held in February 17,
1994 under the aegis of the National Academy's Committee on Astronomy
and Astrophysics and chaired by Al Harper and Anneila Sargent.  Simmons,
G.Rieke, and Werner presented the scientific and technical features of
the warm-launch {\em SIRTF}, which had been adopted as the new baseline
following the recommendation of the SWG.  Its report concluded that
``... {\em SIRTF} remains unparalleled in its potential for addressing
the major questions of modern astrophysics ...'', and we were on our way.

\section{1994 -- 2003}

This began an intensive ~two year period during which the top-level
requirements for the solar-orbit {\em SIRTF} were defined and
reviewed while Jim Fanson at JPL led a design team in fleshing out the
solar-orbit/warm launch concept and JPL, NASA, and Marcia Rieke's
advocacy group pushed for the initiation of the procurement to start the
formal design process.  In the end, we released the RFP for this purpose,
–-- which referred to the ``Green Book'' design completed by Fanson {\em
et al.} but was soliciting team members rather than contractors to build
a point design, in February, 1996.  On June 24, 1996, JPL selected two
major aerospace contractors for three different roles on the team:  Ball
Aerospace was selected to provide the optics, cryogenics, and thermal
shells and shields.  Lockheed-Martin in Sunnyvale was selected to provide
the spacecraft and also for systems engineering and integration and test.
JPL served as the {\em de facto} prime contractor.  Following the
selection of the contractors, Simmons brought key personnel from the
contractors, the instrument teams, JPL, and the SWG together at JPL for
a several-month long period of collocation during which the participating
groups got to know one another and the main interfaces between the various
elements were negotiated.  The various project elements, with the later
addition of the operations elements described below, worked together very
closely and generally amicably up through the launch in 2003.  Along the
way, as detailed by Rieke, we experienced a number of delays and mishaps
attributable to poor performance by the contractors and/or to failure
of oversight by JPL.  However, the effectiveness of this team approach
and the quality of the contractors' work is reflected, in the end,
in the extraordinary performance, reliability, and efficiency of the
{\em Spitzer} observatory on orbit.

\section{The Evolution of {\em Spitzer} Science}

The tight mass and volume constraints of the Delta-launched {\em SIRTF}
led to a scientific payload with far less functionality than envisioned
when the instruments were selected, or for the Titan {\em SIRTF}.
Among the features which have been dropped are narrow field imaging,
polarimetry, and selectable filters for IRAC; spectroscopy shortward of
$5~\mu$m and longward of $40~\mu$m for IRS; and polarimetry, submillimeter
photometry, and selectable filters for MIPS.  However, the instrument
complement now flying is very robust and has the advantages
of no moving parts other than the MIPS scan mirror, and a small number of
operating modes, which simplifies operations.  As the reports elsewhere
in this volume indicate, it also has great scientific power traceable
both to the quality of the detector arrays and to the thoughtful way in
which they are employed within each instrument.

The payload and its functionality are largely unchanged since the CAA
review described above, with one near miss along the way.  In 1994 NASA
began discussions with the Japanese space agency exploring the possibility
of a collaboration on the infrared mission known as {\em IRIS} or {\em
Astro-F}.  {\em IRIS}, now scheduled for launch in 2006, is a $\sim 70$~cm
diameter observatory in LEO which will have both near infrared and far
infrared instruments and operate in both survey and targeted modes.
Dan Weedman encouraged us to pursue this very vigorously, and for a
period of time around in late 1994 there seemed a real possibility
that the IRAC camera, or at least the IRAC functionality, would fly on
{\em IRIS}/{\em Astro-F} and not on {\em SIRTF}.  In the end, however,
it proved impossible to negotiate an acceptable collaboration and IRAC
was welcomed back to {\em SIRTF} in the Spring of 1995.

In 1994, in order to increase the level of community participation and
involvement in {\em SIRTF}, Bob Gehrz, then chair of NASA's Infrared
Management Operations Working Group, became an {\em ex-officio} member
of the {\em SIRTF} SWG.  At the same time, NASA Program Scientist Larry
Caroff initiated discussions aimed at assuring that the {\em SIRTF}
science program was executed in a coherent and well-considered fashion.
The concern was that a program consisting of a number of small projects
might not have maximum lasting impact or archival value.  An additional
concern was the need for early follow-up given the short cryogenic
lifetime anticipated for {\em SIRTF}, which was designed to a 2.5~year
requirement (with a five-year goal).  Over the next few years, a series
of meetings and community workshops under Gehrz' direction produced the
{\em SIRTF} Legacy Program.  Ultimately, six Legacy teams were selected
competitively in late 2000 to carry out large (multi-hundred hour)
projects early in the mission with two provisos:  the pipeline-processed
data would be made public at the same time it was delivered to the
Legacy team, and the Legacy team would contract to provide higher order
data products to the community via the SSC.  This approach to community
engagement in a NASA observatory, particularly with the elimination of
the usual proprietary period, was as much of an experiment as any which
has been carried out with {\em Spitzer}. It was also as successful as
any, and similar programs have subsequently adopted by other missions,
including {\em HST}.  From the {\em Spitzer} perspective, the best
evidence of the success of the Legacy program can be found in the papers
contributed to this volume by the legacy teams and in the large number
of archival research proposals submitted in response to the Cycle 2 call.

Although the hardware elements of the {\em SIRTF} project were well
in place with the selection of the contractor teams in 1996, the
operations planning lagged behind, as is not unusual with NASA missions.
Our shortcomings in the area of science operations were pointed out
rather sharply by Ed Weiler, who became our Program Scientist in 1995
after having served for many years in a similar capacity for {\em HST}.
{\em HST} and {\em Chandra} established the precedent of dedicated
science centers, which carry out a variety of tasks ranging from
soliciting proposals to analyzing and archiving the returned data.
In the case of {\em SIRTF}, it had been tacitly assumed that a similar
center would be formed at or around IPAC, which had been established on
the Caltech campus in 1984 to support the {\em IRAS} data analysis tasks.
With Weiler's encouragement, IPAC was formally named as the home of
the {\em SIRTF} Science Center in July, 1996, and Tom Soifer was named as
the director in 1997.  The SSC (now the {\em Spitzer} Science Center) now
employs about 100 scientists and software engineers.  They work with the
Mission Operations Team at JPL, ably led by Bob Wilson and Chuck Scott,
and the spacecraft operations team based at Lockheed-Martin, Denver, to
keep {\em Spitzer} operating smoothly and with high efficiency.  At the
present time, the observatory is spending about 90\% of wall clock time
executing science and calibration observations, and we have lost only
a few days to spacecraft safing events since launch in August 2003.

\section{The Launch and The Future}

In January, 1996, NASA HQ and JPL signed a Program Commitment Agreement
which formally initiated the {\em SIRTF} Project and projected a launch
date of 2001 December.  The projected cost starting in FY’97 and
including launch costs was \$524M.  In the end, we launched in 2003 August
at a total cost of \$776M.  The cost growth and schedule delays resulted
from a number of causes ranging from problems with the development of
the flight software to concerns about the integrity of the solid fuel
boosters which were strapped to the perimeter of the Delta to provide
added thrust at launch.  Project Managers Larry Simmons and Dave Gallagher
(who replaced Simmons in 1999) steered the project skillfully through this
wide range of obstacles, and the managers at NASA HQ, notably Lia LaPiana,
Ed Weiler and Anne Kinney, were generally supportive of their well-reasoned
requests for additional funds.  The trouble-free operations of {\em
Spitzer} since launch reflect the good use we made of the opportunity
the final delays gave us for operations planning and readiness training.

Following the launch on August 25, 2003, the planned two months of
in-orbit checkout (IOC) activities unfolded very smoothly, thanks to
the excellent work of the planning group led by Sue Linick. Routine
science operations began in early November and were well under way in
mid-December, 2003, when the first data were released and {\em SIRTF} was
officially named for Lyman Spitzer following a naming contest sponsored
by our EPO office.  The winning entry was submitted by Mr. Jay Stidolph
of British Columbia.  The first science publications from {\em Spitzer}
filled a 400+ page edition of the ApJ Supplement on September 1, 2004,
and the scientific triumph of {\em Spitzer} is apparent from the present
volume.

The life-limiting element for {\em Spitzer} seems likely to be the
onboard supply of superfluid liquid helium; as of this writing (March
2005) the spacecraft and instrument warm electronics remain fully
redundant and we have a virtually infinite supply of reaction control gas.
Based on measurements made during IOC and, again, in the Fall of 2004, we
anticipate at least a five-year cryogenic lifetime, extending through the
end of calendar 2008.  The SSC is evaluating techniques for extending the
cryogenic lifetime by up to six months or perhaps even more by carefully
matching the telescope temperature to the exact needs of the instrument
modules in use.  Following cryogen depletion, the telescope and focal
plane should remain cold enough (below $\sim 30$K) to permit use of IRAC
bands 1 and 2 at their full sensitivity.  We can therefore anticipate
scientific use of {\em Spitzer} into the early years of the next decade.

\acknowledgements 

The {\em Spitzer Space Telescope} is operated by the Jet Propulsion
Laboratory, California Institute of Technology, under NASA Contract 1407.
The opinions expressed in this article are entirely the author's,
not those of JPL/Caltech or NASA.  The author thanks George Rieke for
comments and for sharing a {\em Spitzer} chronology, Larry Simmons for
comments, and Daniel Stern for assistance with preparing the manuscript.


\end{document}